\documentclass[twocolumn,letterpaper,letter]{jpsj2}

\makeatletter
\def\@cite#1{\textsuperscript{#1)}}
\makeatother
\title{Knudsen Effect in a Nonequilibrium Gas}
\author{
Taka H. Nishino and Hisao Hayakawa }
\inst{Department of Physics at Yoshida-South Campus, Kyoto University, Sakyo-ku, Kyoto 606-9501, JAPAN.}
\abst{
 From the molecular dynamics simulation of a system of hard-core disks
 in which an equilibrium cell is connected with a nonequilibrium cell,
 it is confirmed that the pressure difference between two cells depends
 on the direction of the heat flux.
 From the boundary layer analysis,
 the velocity distribution function in the boundary layer is obtained.
 The agreement between the theoretical result and the numerical result
 is fairly good.
}
\kword{molecular dynamics simulation, nonequilibrium steady state,
heat conduction, boundary layer analysis,
Boltzmann equation, Knudsen effect}
\begin{document}
\maketitle
Although there has been a long history of nonequilibrium statistical mechanics
since Boltzmann introduced the Boltzmann equation,
the understanding of nonequilibrium statistical mechanics
is still in the primitive stages \cite{cc,sone2,pk}.
The significant role of nonequilibrium physics in the
mesoscopic region has been recently recognized.
For example,
there has been some important progress such as
the Fluctuation Theorem \cite{ft,ft2} and the Jarzynski
equality \cite{je} in mesoscopic nonequilibrium statistical mechanics.
In typical situations of mesoscopic physics, materials are confined to
narrow regions.
Thus the boundary effects at the mesoscopic scale are important not only for
nonequilibrium statistical mechanics but also for
consideration of friction and lubrication~\cite{pt}.

On the other hand, some macroscopic phenomenologies for 
nonequilibrium steady states have been proposed.
These are the Extended Thermodynamics (ET) \cite{et},
the Extended Irreversible Thermodynamics (EIT) \cite{eit,eit2,eit3} which
is the combination of ET and information theory, and
the Steady State Thermodynamics (SST) \cite{sst}.
It is interesting that both EIT and SST treat a common process in
which a nonequilibrium cell is connected with an equilibrium cell
\cite{sst,eit2}.

The Knudsen effect is the phenomenon in which two
equilibrium cells with different temperatures are connected by a small
hole \cite{pk}.
The balance equation of the Knudsen effect is given by
\begin{equation}
 \label{knudsen}
 \frac{P_1}{\sqrt{T_1}} = \frac{P_2}{\sqrt{T_2}},
\end{equation}
where $T_i$ and $P_i$ represent the temperature and the pressure in the cell,
$i$, respectively.
Although eq. (\ref{knudsen}) contains only
macroscopic variables,
it is in contrast to the ordinary thermodynamic balance
condition where the pressures the two cells are equal.
Such an exceptional condition means that
the mass balance is determined by the transportation of the
gases in the small hole.
Therefore,
the relevance of predictions by macroscopic theory \cite{eit3,sst}
for the generalized Knudsen effect in
which an equilibrium cell is connected with a nonequilibrium cell by a
small hole is questionable.

On the other hand,
the explicit perturbative solution of the Boltzmann equation for hard
spheres has been 
derived at the Burnett order of the heat flux \cite{kim-hayakawa,kim}.
The quantitative accuracy and numerical stability of their solution in
the bulk region has
been confirmed by molecular dynamics (MD) simulation for hard-spheres
\cite{fushiki} and the extension to the tenth order shows that their
second order solution is accurate even when the heat current is
large \cite{fushiki2}.
They have confirmed that the solution in the bulk region
derived by information theory
is not consistent with that of the Boltzmann
equation \cite{kim-hayakawa2}.
Kim and Hayakawa \cite{kim-hayakawa} 
have also discussed the nonequilibrium Knudsen effect.
Their result denies the prediction of SST,
but both theories predict that the osmosis
$\triangle P$ defined by the $\triangle P \equiv P_{xx}^{neq}-P^{eq}$ with
the $xx$ component of the pressure tensor $P_{xx}^{neq}$ in
the nonequilibrium cell and the pressure
$P^{eq}$ in the equilibrium cell
is always positive regardless of the
direction of heat flux.

However, their simplification \cite{kim-hayakawa}
using the bulk
solution of the Boltzmann equation in the boundary layer is not
acceptable.
In fact, numerical simulations of the Boltzmann
equation of hard spheres \cite{sone} exhibit the discontinuity of velocity
distribution function (VDF) in the boundary layer, but there is no
discontinuity of VDF derived by Kim and Hayakawa \cite{kim-hayakawa,kim}.
Furthermore, it is also well 
known that the gas temperature near the wall is different from the
wall temperature \cite{sone,pk,sone2,welander},
but both treatments \cite{sst,kim-hayakawa} ignore this fact.

\begin{figure}
 \centering
 \includegraphics[width=8.5cm]{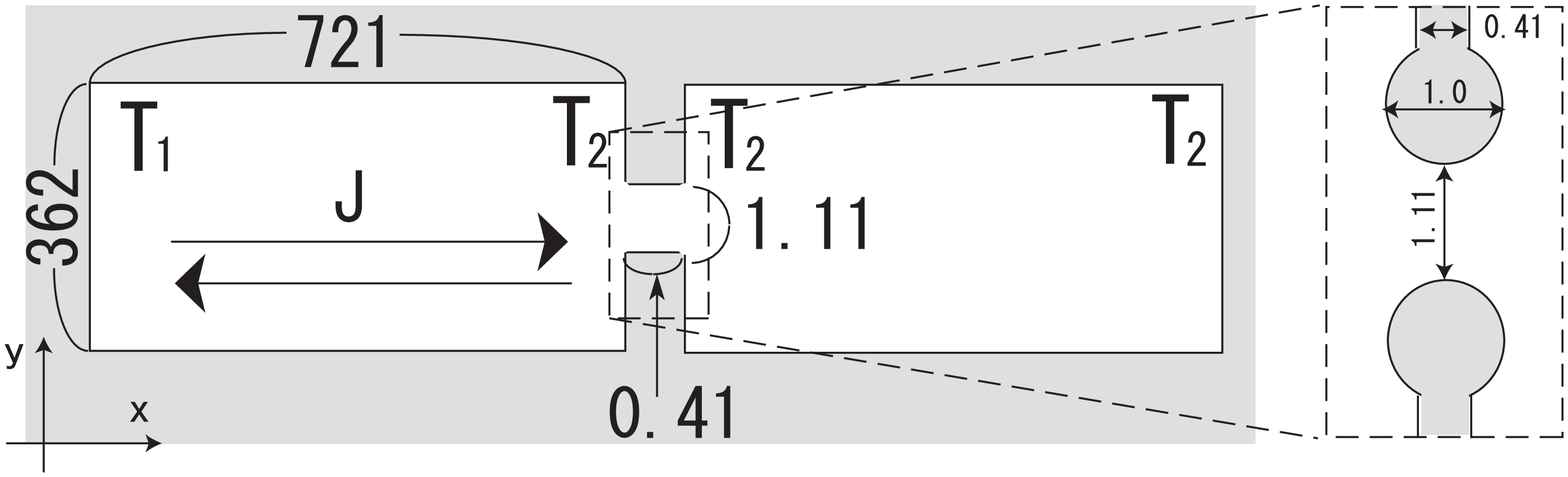}
 \caption{
 The schematic picture of the system of our simulation, where 
 the right cell is at equilibrium with temperature
 $T_{2}$ and the left cell is in a nonequilibrium state.
 The inset represents the detailed structure of the hole that has the
 gap $1.11$.
 The unit of length in figures is the diameter of the rigid disk.
 We adopt periodic boundary condition in the $y$ direction. }
 \label{neq}
\end{figure}

To clarify the truth of the nonequilibrium Knudsen effect,
we employ the event-driven molecular dynamics simulation
of hard-disks developed in refs. [20-23].
Let $x$ and $y$ be the Cartesian coordinates of the horizontal and the
vertical directions, respectively.
First we check the validity of eq. (\ref{knudsen}) by MD.
We adopt the diffusive reflection for the
vertical walls away from the
hole, the simple reflection rule for the
wall between two cells and the periodic boundary condition
for the horizontal wall.
We connect two cells by a small hole, as illustrated in Fig. \ref{neq}.
For $T_1/T_2 = 1.96$, we have found the steady value of
$P_1/P_2 \sqrt{T_2/T_1} = 0.9982 \pm 0.0195$,
where the average area fraction and the number of the particle
are $0.015$ and $10,000$ respectively.
We determine the pressure based on the Virial
theorem \cite{hansen} and average in $10^4$ collisions per particle.
The stationary state is realized after
$3 \times 10^5$ collisions per particle and we
use the data after that.
We simulate the system until $10^6$ collisions have been
performed per particle.

Next, we simulate the nonequilibrium Knudsen effect.
The number of hard disks and the average area fraction
are the same values used to simulate the conventional Knudsen effect.
We adopt the diffusive boundary condition for the wall between two cells
(see Fig. \ref{neq}).
We examine the values of $T_1/T_2$ as $1.96$ and $1/1.96$.
We cause both cells to divide into 10 equal parts, and
we have confirmed that pressure based on the Virial
theorem \cite{hansen} is identical in each cell.
This is consistent
with the stationary condition \cite{welander}.
In nonequilibrium cases the stationary state is realized after
$10^6$ collisions per particle and we use the data
after that.
We simulate the system until $4.4 \times 10^6$ collisions have been performed
per particle.

\begin{figure}
 \centering
 \includegraphics[width=8.5cm]{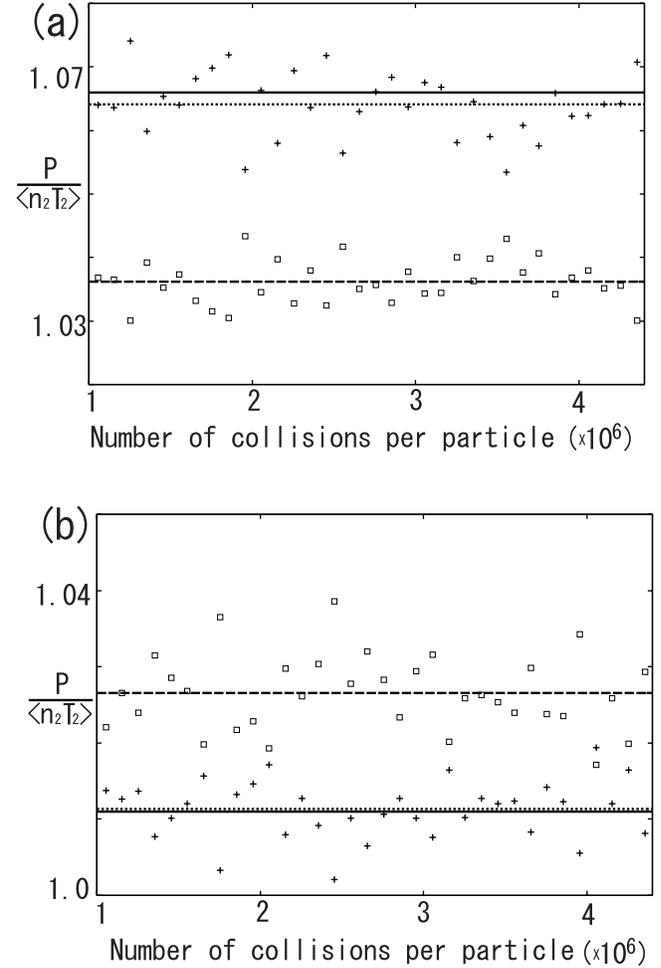}
 \caption{
 The time dependence of the pressure for (a) $T_1/T_2 = 1.96$ and (b)
 $T_1/T_2 = 1/1.96$.
 Here $P^{neq}_{xx}$ in the nonequilibrium cell is plotted
 by plus points and its time average is shown by the dotted line.
 In each case
 the pressure $P^{eq}$ in the equilibrium cell is plotted
 by open square points
 and its time average is plotted by the dashed line.
 The theoretical $P^{neq}_{xx}$ in eq.
 (\ref{yosou}) is shown by the solid line.
 For the normalization, we divide these values by $<n_2T_2>$.
 }
 \label{pressure}
\end{figure}

Our results of MD plotted in Fig. \ref{pressure} indicate
that the sign of $\triangle P (\equiv P^{neq}_{xx} - P^{eq})$ depends on the
direction of energy flux.
Actually, the stationary value $\triangle P$ in our simulation
is given by $\triangle P/<n_2T_{2}> = 0.0279 \pm 0.0184$ for
$T_{1}/T_{2} = 1.96$, and
$\triangle P/<n_2T_{2}> = -0.0152 \pm 0.0175$ for
$T_{1}/T_{2} = 1/1.96$, where $<n_2T_2>$ is the time average of $n_2T_2$
(Fig. \ref{pressure}).
This result contrasts with the positive  $\triangle P$
predicted by both SST \cite{sst} and Kim and Hayakawa \cite{kim-hayakawa}.

\begin{figure}
 \includegraphics[width=8.5cm]{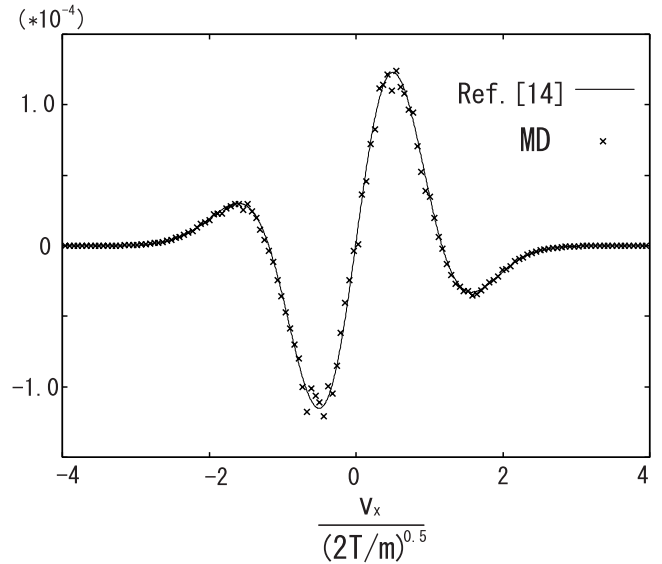}
 \caption{\label{middle}
 Comparison of VDFs
 derived by simulation and Ref.[14] at the 
 middle of the nonequilibrium cell.
 The plotted data are the substraction of Maxwellian
 from VDFs.
 }
\end{figure}

\begin{figure}
 \includegraphics[width=8.5cm]{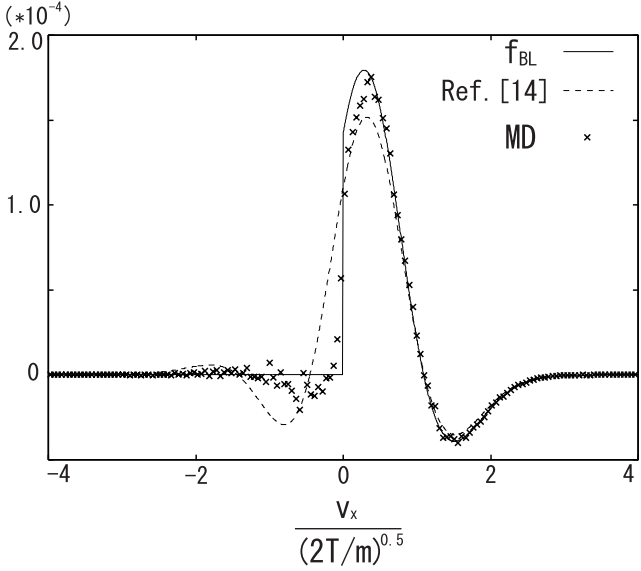}
 \caption{\label{hashi}
 Comparison of VDFs derived by
 simulation and Ref.[14] and
 $f_{BL}$ in eq. (\ref{function}) at the near the wall.
 The plotted data are the substraction of Maxwellian
 from VDFs.}
\end{figure}

Now let us compare the
VDF of MD with the perturbative solution of the Boltzmann
equation at the Burnett order obtained by Kim \cite{kim} for 2D
hard disks.
From now on, we restrict our interest to the data for heat flux
$J<0 $, with $T_1/T_2 = 1/1.96$.
As shown in Fig. \ref{middle}, VDF obtained from MD in the bulk
region of a nonequilibrium cell is almost identical to the theoretical
VDF, where the VDF of MD is obtained from the average of particles existing in
$\pm1.05d$ from the center of nonequilibrium cell in the horizontal
direction with hard disk diameter $d$.
On the other hand, the VDF
in the boundary layer in which we average the data of particles existing
between $0.7d$ and $2.8d$ apart from the right wall of
the nonequilibrium cell deviates from the theoretical VDF \cite{kim},
particularly for $v_x < 0$ (Fig. \ref{hashi}).

Let us derive VDF in the boundary layer in the
nonequilibrium cell.
We assume that the VDF of the
incident particles ($v_x > 0$) in the boundary layer
is the same as the bulk distribution function ($f_{NE}$).
On the other hand, we assume that the VDF for
the particles reflected by the wall
($v_x < 0$) obey the Maxwellian.

Thus, the distribution function $f_{BL}$ in the boundary layer is given by
\begin{equation}
 \label{function}
  f_{BL}({\bf v}) = \begin{cases}
      f_{MB}(n_w,T_w,{\bf v}) & v_x < 0 \\
      f_{NE}(n_x,T_x,J,{\bf v}) & v_x \ge 0 ,
		    \end{cases}
\end{equation}
where
\begin{equation}
 \label{fmb}
 f_{MB}(n_w,T_w,{\bf v}) \equiv \frac{n_w m}{2 \pi T_w} \exp{\left[-\frac{m {\bf v}^2}{2T_w}\right]},
\end{equation}
with the mass of a hard-disk $m$, and
\begin{equation}
 \label{fbl}
  \begin{split}
 f_{NE}(n_x,T_x,J,&{\bf v}) \equiv f_{MB}(n_x,T_x,{\bf v}) \times \\
  &\left[1-\frac{m J v_x}{2 b_1 n_x T_x^2} \sum_{r\ge1} r!b_rL^1_r\left(\frac{m {\bf v}^2}{2T_x}\right)\right],
   \end{split}
\end{equation}
with $b_1=1.03, b_2=5.738 \times 10^{-2}, b_3=4.946 \times 10^{-3}, b_4
= 4.313 \times 10^{-4}, b_5 = 3.452 \times 10^{-5}, b_6 = 2.241 \times
10^{-6}$
and Laguerre's bi-polynomial $L^a_b (x)$ \cite{kim}.
Since the heat flux is sufficiently small, we adopt the first order
nonequilibrium VDF in the heat-flux for $f_{NE}$.

There are three unknown variables, $n_w$, $n_x$, and $T_x$ in
eqs. (\ref{fmb}) and (\ref{fbl}), while there are three relations,
\begin{align}
 \label{ng}
 n & \equiv \int d{\bf v} \, f_{KL}({\bf v}), \\
 \label{Jg}
 J & \equiv m \int d{\bf v} \, v_x \frac{{\bf v}^2}{2} f_{KL}({\bf v}) ,\\
  \label{massflux}
 & \int d{\bf v} \, v_x f_{KL}({\bf v}) = 0.
\end{align}
Here, the first two equations are definitions of the density
$n$ and the heat flux $J$, and
the last equation represents the mass balance condition.
Therefore, we can determine $n_w$, $n_x$, and $T_x$ from
eqs. (\ref{ng})-(\ref{massflux}).
The expansions in terms of $J$ of the three variables become
\begin{align}
 \label{nw}
 &n_w = n + a \frac{m^{1/2}}{ T_w^{3/2}} J,\\
 \label{nx}
 &n_x = n + b \frac{m^{1/2}}{ T_w^{3/2}} J,\\
 \label{Tx}
 &T_x = T_w + c \frac{m^{1/2}}{ n T_w^{1/2}} J,
\end{align}
where $a$, $b$ and $c$ are constants to be determined.
From eqs. (\ref{ng}),(\ref{Jg}) and (\ref{massflux}), we obtain the 
solutions of the linear simultaneous equations
as $(a,b,c) \simeq (0.32,-0.099,0.84)$.

The distribution function $f_{BL}$ near the wall has thereby been
determined by $n, T_w$, and $J$.
Figure \ref{hashi} is the comparison of $f_{BL}$ with the
results of MD.
From Fig. \ref{hashi}, both $f_{BL}$ and the VDF from MD have a
discontinuity at $v_x = 0$, as in the case of conventional boundary
layer analysis \cite{sone}.
For reflective VDF ($v_x < 0$), there is
the small difference between the result of MD and the Maxwellian.
We may deduce that it arises from collisions between particles because
we measure the VDF a short distance from the wall.

With the aid of $f_{BL}$, $\triangle P$ can be calculated as
\begin{equation}
 \label{yosou}
 \triangle P
  \simeq \frac{-a+b+c+0.412}{2}\frac{J}{\sqrt{T_{2}}}
  = 0.415 \frac{J}{\sqrt{T_{2}}}.
\end{equation}
From this equation and the value of the heat-flux in MD, we evaluate
$\triangle P/<n_{2}T_{2}> \simeq 0.0298$ for $T_{1}/T_{2} = 1.96$ and
$\triangle P/<n_{2}T_{2}> \simeq -0.0156$ for $T_{1}/T_{2} = 1/1.96$.
The result agrees well that of the simulation.
We can also obtain the temperature jump coefficient
$\gamma \simeq 0.97$ which is defined through 
$T_{x=0} - T_w = \gamma J/n \sqrt{m/T_w}$.

Now, let us discuss our result.
There are some advantages to employing MD as the numerical simulation.
First, we can easily change boundary conditions for walls
depending on our interest.
The system of connecting two cells by a small hole is easily simulated by MD.
Second, MD is suitable for high density simulation of gases.
The nonequilibrium Knudsen effect for high density gases
is an interesting subject for future discussions.
We are also interested in the size effect
of the hole on the transition from Knudsen
balance to the thermodynamic balance.
On the other hand, there are some disadvantages to
employing MD.
Because of the small system size of our MD simulation,
the fluctuation of the pressure is large.

The advantage of our boundary layer analysis is that we can
write the explicit form
of VDF in the boundary layer in terms of the density, the heat-flux
and the temperature of the wall.
On the other hand, VDF by Sone \textit{et al} \cite{sone} has an implicit
form that is obtained as a numerical solution of an integral equation.
The explicit VDF can be obtained by our simplification, but the validity
of this method has not been confirmed.
In fact, for 3D hard-sphere gases, our method predicts the
temperature jump coefficient $\gamma \simeq 0.72$, while
Sone \textit{et al} \cite{sone} predicts $\gamma  \simeq 1.00$.
To check the validity of our boundary layer analysis,
it will be necessary to employ 3D MD simulation for 3D.

We also stress the difficulty of describing
the nonequilibrium Knudsen effect by macroscopic phenomenology.
In fact, our boundary layer analysis
strongly depends on the boundary condition.

In conclusion, the sign of
$\triangle P$ depends on the direction of the heat flux.
The approximated VDF in the boundary layer
has been obtained from the assumption
that the incident particles obey the bulk VDF and the
reflected particles obey the Maxwellian.
This agrees well with the simulation result.

The authors would like to thank M. Fushiki and H.-D. Kim for their
valuable comments.
They appreciate Aiguo Xu for his critical reading of the manuscript.
This work is partially supported by a Grant-in-Aid from the Japan Space
Forum and the Ministry of Education, Culture, Sports, Science and
Technology (MEXT), Japan (Grant No. 15540393)
and a Grant-in-Aid from the $21$st century COE ``Center for Diversity
and Universality in Physics'' from MEXT, Japan.

\end{document}